\documentclass [12 pt]{article}
\begin{document}
\title{Estimating  Net Effects of Treatments in Treatment Sequence without the Assumption of Strongly Ignorable Treatment Assignment}
\author{Li Yin  \& Xiaoqin Wang}
\maketitle                   

\noindent
Li Yin   is a senior researcher at  Karolinska Institute, Box 281, SE
171 77, Stockholm, Sweden (Email: li.yin@ki.se). Xiaoqin Wang is a
senior lecturer at  University of G\"{a}vle, 801 76, G\"{a}vle, Sweden
(Email: xwg@hig.se).

\date{}

\begin{abstract}
In sequential causal inference, one  estimates the causal net effect of treatment in  treatment sequence on an outcome after last treatment in the presence of   time-dependent covariates  between treatments,    improves the estimation by the untestable assumption of strongly ignorable treatment assignment, and  obtains consistent but non-genuine likelihood-based estimate.
In this article, we introduce the net effect of  treatment  as  parameter for the conditional distribution of outcome  given all   treatments and  time-dependent covariates and show that it is equal to the causal net effect of treatment under the assumption of strongly ignorable treatment assignment.  As a result, we can estimate the net effect of treatment  and evaluate its causal interpretation  in two separate steps. The first step is fucus of this article while the second step can be accomplished by  usual sensitivity analyses.
We construct
point parametrization for the conditional   outcome distribution in which the parameters of interest are the point effects of single-point treatments. With  point parametrization and without the untestable assumption, we estimate the net effect of treatment by maximum likelihood, improve the estimation by testable pattern of the net effect of treatment, and obtain
unbiased consistent maximum-likelihood estimate for the net effect of treatment with  finite-dimensional  pattern.
\end{abstract}





\section{Introduction}
In many economic and medical  practices, a sequence of treatments, i.e. economic interventions or medical treatments or exposures,  are assigned  to influence an outcome of interest that occurs after last treatment of the sequence. Between consecutive treatments,    time-dependent covariates are present that   may be posttreatment variables of the earlier treatments (Rosenbaum, 1984;
Robins, 1989; Frangakis \& Rubin, 2002) and confounders of the subsequent  treatments.
Under the assumption of strongly ignorable treatment assignment or called the assumption of no unmeasured confounders,
Robins (1986, 19992, 1997, 1999, 2004, 2009) identified  the causal net effect of each treatment in treatment sequence by standard parameters, which are usually the means of    outcome given   all treatments and   time-dependent covariates. The causal net effect of treatment
is also called the blip effect of treatment in the context of semi-parametric sequential causal inference.

Robins   illustrated that  any constraint   imposing equalities
among  standard parameters   leads to erroneous rejection
of the null hypothesis of causal net effects of treatments  if the  time-dependent covariates are
simultaneously  posttreatment variables of the earlier treatments and confounders  of the subsequent  treatments.
As treatment sequence gets long, the number of standard parameters becomes huge, and with no constraint on these parameters, the maximum-likelihood (ML)  estimates of causal net effects  of treatments may not be consistent (Robins \& Ritov, 1997; Robins, 1997).

To overcome this difficulty, two semi parametric approaches have been developed, the $g$-estimation   model (Robins, 1992,
1997, 2004, 2009; Robins et al., 1999; Henderson et al., 2010) and the  marginal structural model (Robins, 1999, 2009;
Murphy et al., 2001). The  $g$-estimation   model  uses the assumption of    strongly ignorable treatment assignment and
  is based on the likelihood of treatments given   previous time-dependent covariates and treatments. The  marginal structural model also uses the assumption and is based on a weighted conditional likelihood of outcome given all   treatments and time-dependent covariates.
Both approaches yield consistent but non-genuine likelihood-based estimates of causal net effects of treatments. Both approaches are dependent on validity of the assumption, which is noticeably untestable with observed  data.

In this article, we intend to estimate the causal net effect of treatment by maximum likelihood and improve the estimation by testable assumptions.
To this end,  we
introduce the net effect of   treatment    as parameter for the conditional distribution of outcome given all treatments and time-dependent covariates and show that the parameter  is equal to the causal net effect of treatment under the assumption of strongly ignorable treatment assignment. As a result, we can   estimate the net effect of treatment and  evaluate its causal interpretation  in two separate steps. The first step is focus of this article whereas the second step can be accomplished by using subject knowledge in combination of usual sensitivity analyses.  We use testable pattern of the net effect of treatment to improve the estimation and obtain
 unbiased consistent ML estimate of the net effect of treatment if the pattern is of finite dimension.

In Section $2$, we describe the relationship between the causal net effect of treatment and the net effect of treatment. In Section $3$,
we construct  point
parametrization for the conditional  outcome distribution by using   point effects of treatments or time-dependent covariates as point parameters and express pattern of net effects of treatments  by constraint on point effects of treatments.
In Section $4$, we estimate net effects of treatments  by maximum likelihood in the point parametrization  and show that the ML estimates are both unbiased and consistent in  many practical applications where the net effects of treatments have pattern of finite dimension.
In Section $5$, we study an example in which an employer rewards a sequence of bonuses to the employees in order to increase their productivity,  but wonders if it is possible to reward bonuses less frequently while not reducing the productivity, and in Section $7$, we continue the example to illustrate practical procedure of estimating net effects of treatments.
In Section $7$, we conclude the article
with remarks.

\newtheorem{E}{Example}
\newtheorem{T}{Theorem}
\newtheorem{PP}{Proposition}
\newtheorem{C}{Corollary}
\newtheorem{D}{Definition}
\newtheorem{R}{Remark}
\newtheorem{LL}{Lemma}

\section{Net Effects versus Causal  Net Effects of Treatments in Treatment Sequence}

\subsection{Treatment sequence, time-dependent covariates and  outcome}

Let  $z_t$ indicate the treatments at time $t$ ($t=1,\ldots,T$). Assume that all  $z_t$ are   discrete variables  and take the values $0,1,\ldots$.
We take $z_t=0$ as control treatment and $z_t=1,2,\ldots$ as active treatments. Let
$\mathbf{z}_1^t=(z_1,\ldots, z_t)$.
 For notational simplicity, we use
 one subpopulation defined  by   stationary covariates of the population as our population,  and henceforth do not consider  stationary covariates in the following development.

Between treatments
$z_t$ and $z_{t+1}$ ($t=1,\ldots,T-1$), there is a time-dependent
covariate vector $\mathbf{x}_t$, which can be confounders for subsequent treatments $z_s$ ($s=t+1,\ldots, T$) and posttreatment variables of earlier treatment  $z_s$ ($s=1,\ldots, t$).
Assume that $\mathbf x_t$ is a discrete  vector with non-negative components. We take $\mathbf x_t=\mathbf 0$ as  reference level.
Let
$\mathbf{x}_{1}^{t}$ $=$ $\{\mathbf{x}_{1},
\mathbf{x}_{2},\ldots,
\mathbf{x}_{t}\}$  be the  time-dependent
covariate array between treatments $z_1$ and $z_{t+1}$. The outcome of interest after last treatment $z_T$ is denoted by $y$.

Instead of one set $(\mathbf z_{1}^T, \mathbf{x}_{1}^{T-1},y )$ of the random  variables,  we consider
 $N$ independent and identically distributed  sets,
 $\{
\mathbf z_{i1}^T, \mathbf{x}_{i1}^{T-1}, y_i\}$, $i=1,\ldots, N$. In this article, we shall ignore the variability of $\{
\mathbf z_{i1}^T, \mathbf{x}_{i1}^{T-1}\}_{i=1}^N$ and  focus on the conditional distribution of $\{y_i\}_{i=1}^N $ given $\{(\mathbf z_{i1}^T, \mathbf{x}_{i1}^{T-1})\}_{i=1}^N $, that is,
\begin{equation}\label{eq1}
\prod_{i=1}^N f(y_i\mid \mathbf z_{i1}^T, \mathbf{x}_{i1}^{T-1} ).
\end{equation}
 Noticeably, standard parameters for (\ref{eq1}) are the means $\mu(\mathbf z_1^{T},\mathbf {x}_1^{T-1})=E(y \mid
  \mathbf z_1^{T},\mathbf {x}_1^{T-1})$,  where the expectation  $E(b\mid a)$ is   with respect to the conditional distribution of   $b$   given $a$.

Throughout the article, we adopt the following  notational conventions. First, the notations
$\mathbf z_{u}^{v}$ and $\mathbf x_{u}^{v}$  with $u > v$ or $u=v=0$ or both $u<0$ and $v<0$  should be omitted from    relevant
expression.
For instance,  the notation  $\mathbf x_1^{0}$
in $\mu(\mathbf z_1^{T},\mathbf {x}_1^{T-1})$ for $T=1$ should be omitted, so that we have $\mu(z_1)$.   Second, the sigma notation  $\sum_{i=u}^v a_i$ with $v<u$ should be omitted from   relevant expression.
Third, the notations $\mathbf z_{u}^{v}$, $\mathbf x_{u}^{v}$  and $\sum_{i=u}^v a_i$  with $u < 1 $ and $v \geq 1$ are treated  as $\mathbf z_{1}^{v}$ and $\sum_{i=1}^v a_i$ respectively. Fourth, the notation $(\mathbf z_u^{v},\mathbf x_u^{v-1})$ is equal to $(\mathbf z_u^{v-1}, \mathbf {x}_u^{v-1}, z_v)$, and $(\mathbf z_u^{v},\mathbf x_u^{v})$ to $(\mathbf z_u^{v}, \mathbf {x}_u^{v-1}, \mathbf x_v)$; we may use one  or another notation in different contexts.

\subsection{Net   effects and causal net effects of treatments}
Given   $N$ sets $\{\mathbf z_{i1}^T, \mathbf x_{i1}^{T-1}\}_{i=1}^N$, a stratum is a set of those   sets   satisfying certain condition. For instance, stratum $(\mathbf z_1^{t}, \mathbf x_1^{t-1})$ is a set of those   sets    satisfying
$(\mathbf z_{i1}^{t}, \mathbf x_{i1}^{t-1})=(\mathbf z_1^{t}, \mathbf x_1^{t-1})$.
Let   ${\rm pr}(A)$ denote the proportion of stratum $A$ in the $N$ sets and ${\rm pr}(A\mid B)$ denote the conditional proportion of stratum $A$ in stratum $B$.

We define the net effect of last treatment $z_T>0$ on stratum $(\mathbf z_1^{T-1}, \mathbf x_1^{T-1})$  by
$$
 \phi(\mathbf z_1^{T-1}, \mathbf x_1^{T-1}, z_T)=\mu(\mathbf z_1^{T-1}, \mathbf x_1^{T-1}, z_T)-\mu(\mathbf z_1^{T-1}, \mathbf x_1^{T-1}, z_T=0).
$$
The mean of $y$ in stratum $(\mathbf z_1^{T-2}, \mathbf x_1^{T-2}, z_{T-1})$ is
$$
  \mu(\mathbf z_1^{T-2}, \mathbf x_1^{T-2}, z_{T-1})= \sum_{
\mathbf
x_{T-1}, z_T} \mu( \mathbf z_1^{T},\mathbf
x_1^{T-1}) {\rm pr}(\mathbf x_{T-1}, z_T\mid
\mathbf z_1^{T-2},\mathbf x_1^{T-2}, z_{T-1}).
 $$
With $ \phi(\mathbf z_1^{T-1}, \mathbf x_1^{T-1}, z_T)$, we calculate
 the mean of $y$ in stratum $(\mathbf
z_1^{T-2},\mathbf x_1^{T-2}, z_{T-1})$ under no active treatments at $t=T$  by
$$
\nu(\mathbf z_1^{T-2}, \mathbf x_1^{T-2}, z_{T-1})=
$$
$$\mu(\mathbf z_1^{T-2}, \mathbf x_1^{T-2}, z_{T-1})- \sum_{
\mathbf x_{T-1}}\sum_{z_T>0  }\phi(\mathbf z_1^{T-1}, \mathbf
x_1^{T-1},z_T){\rm pr}(
\mathbf x_{T-1},z_T\mid \mathbf
z_1^{T-2},\mathbf x_1^{T-2}, z_{T-1}).
$$
With $\nu(\mathbf z_1^{T-2}, \mathbf x_1^{T-2}, z_{T-1})$,  we define the net effect of second last treatment $z_{T-1}>0$ on stratum $(\mathbf z_1^{T-2}, \mathbf x_1^{T-2})$  by
$$
 \phi(\mathbf z_1^{T-2}, \mathbf x_1^{T-2}, z_{T-1})=\nu(\mathbf z_1^{T-2}, \mathbf x_1^{T-2}, z_{T-1})-\nu(\mathbf z_1^{T-2}, \mathbf x_1^{T-2}, z_{T-1}=0).
$$

Recursively, we define  the net effect $\phi(\mathbf z_1^{t-1}, \mathbf x_1^{t-1}, z_{t})$  of  treatment $z_t>0$ on stratum $(\mathbf z_1^{t-1}, \mathbf x_1^{t-1})$ ($t=T-2, \ldots, 1$).
In summary, we have, for $t=T,\ldots, 1$,
$$
\mu(\mathbf z_1^{t-1}, \mathbf x_1^{t-1}, z_t)= \sum_{\mathbf z_{t+1}^{T},
\mathbf
x_{t}^{T-1}} \mu( \mathbf z_1^{T},\mathbf
x_1^{T-1}) {\rm pr}(\mathbf z_{t+1}^{T},\mathbf x_{t}^{T-1}\mid
\mathbf z_1^{t},\mathbf x_1^{t-1}),
$$
$$
\nu(\mathbf z_1^{t-1}, \mathbf x_1^{t-1}, z_{t})=
$$
$$\mu(\mathbf z_1^{t-1}, \mathbf x_1^{t-1}, z_{t})- \sum_{s=t+1}^T\sum_{
\mathbf z_{t+1}^{s-1},\mathbf x_{t}^{s-1}}\sum_{z_s>0  }\phi(\mathbf z_1^{s-1}, \mathbf
x_1^{s-1},z_s){\rm pr}(
\mathbf z_{t+1}^{s-1},\mathbf x_{t}^{s-1},z_s\mid \mathbf
z_1^{t},\mathbf x_1^{t-1}),
$$
\begin{equation}\label{eq2}
 \phi(\mathbf z_1^{t-1}, \mathbf x_1^{t-1}, z_{t})=\nu(\mathbf z_1^{t-1}, \mathbf x_1^{t-1}, z_{t})-\nu(\mathbf z_1^{t-1}, \mathbf x_1^{t-1}, z_{t}=0).
\end{equation}
Noticeably,  $\nu(\mathbf z_1^{T-1}, \mathbf x_1^{T-1}, z_{T})=\mu(\mathbf z_1^{T-1}, \mathbf x_1^{T-1}, z_{T})$, according to the notational convention described in Section $2.1$.
Given $\{
\mathbf z_{i1}^T, \mathbf{x}_{i1}^{T-1}\}_{i=1}^N$, the proportions of treatments and covariates can be treated as constants. Therefore
the net effects   are linear functions of the standard parameters $\mu(\mathbf z_1^{T-1}, \mathbf x_1^{T-1}, z_{T})$ and thus
 are  parameters of the  outcome distribution (\ref{eq1}). These parameters evaluate the association between treatment $z_t$ ($t=1,\ldots, T$) and the outcome $y$.

Let $\mathbf z_t^{T}=(z_t,\ldots, z_{T})$ be the treatment sequence  given the  variables $(\mathbf z_1^{t-1}, \mathbf x_1^{t-1})$. Under  $\mathbf
z_t^{T}$ given $(\mathbf z_1^{t-1}, \mathbf x_1^{t-1})$, each unit could have  a potential outcome $y(\mathbf z_1^T)$. Let  $y(\mathbf z_t^{T})=y(\mathbf z_1^{T})$ for given $(\mathbf z_1^{t-1}, \mathbf x_1^{t-1})$.
Under the assumption of strongly ignorable treatment assignment (Robins, 1986, 1989, 1992, 1997, 1999, 2004, 2009), we show, in Appendix $A1$,
$$
\nu( \mathbf
z_1^{t-1},\mathbf{x}_1^{t-1},z_t) = E\{y( z_t, \mathbf z_{t+1}^{T}=\mathbf 0)\mid
  \mathbf z_1^{t-1},\mathbf{x}_1^{t-1}\},
$$
$$
 \phi(\mathbf z_1^{t-1}, \mathbf x_1^{t-1}, z_{t})=
$$
\begin{equation}\label{eq4}
 E\{y( z_t, \mathbf z_{t+1}^{T}=\mathbf 0)\mid
  \mathbf z_1^{t-1},\mathbf{x}_1^{t-1}\}- E\{y( z_t =0, \mathbf
z_{t+1}^{T}=\mathbf 0)\mid   \mathbf z_1^{t-1},\mathbf{x}_1^{t-1}\}
\end{equation}
for $t=1,\ldots, T$, which is the causal net effect of treatment $z_t > 0$ on stratum  $ (\mathbf z_1^{t-1},\mathbf{x}_1^{t-1})$  (Robins, 1986, 1989, 1992, 1997, 1999, 2004, 2009).

\subsection{Difficulties in estimation of  net effects of treatments in standard parametrization}
If $\mathbf x_t$ are   posttreatment variables of   $z_s$ ($s \leq t$), then  the standard parameters $\mu(\mathbf z_1^{T},\mathbf x_1^{T-1})$ essentially do not have any pattern (Rosenbaum, 1984;
Robins, 1989; Frangakis \& Rubin, 2002). If $ \mathbf x_t$  are  simultaneously confounders of   $z_s$  ($s > t $), then one needs to use all  the standard parameters to identify the  causal net effects of treatments (Robins, 1986, 1997, 1999, 2004, 2009; Robins \& Ritov, 1997).
 As $T$ increases, the number of standard parameters increases exponentially and the ML  estimates of  the causal net effects may not be consistent (Robins \& Ritov, 1997).  In general, the difficulty   applies to the net effects.

Although standard parameters do no have pattern, the net effects may have one, which is the focus of this article. Pattern of net effects implies
constraint on standard parameters.
Consider a simple case where $z_t=0,1$ ($t=1,\ldots, T-1$) and $x_t=0,1$ ($t=1,\ldots, T-1$). Then there are as many as $2^{2T-1}$ standard parameters. Suppose that the pattern of net effects is such that  all the net effects   are the same, denoted by $\varphi$. Then there exist as many as   $(4^T-1)/(4-1)-1$ equalities among   standard parameters. For $T=10$, we may have to solve a system of $524288$ likelihood equations under a constraint of $349524$ equalities to estimate  $\mu(\mathbf z_1^{T},\mathbf x_1^{T-1})$ and then $\varphi$. As $T$ increases, the number of likelihood equations  increases exponentially and so does the number of equalities among standard parameters, and it is practically impossible to  solve such a huge system  of  likelihood equations under a constraint of so many equalities.

\section{Point versus Net Effects of Treatments in Treatment Sequence}

\subsection{Point effects of treatments and point  parametrization}
The mean of $y$ in stratum $(\mathbf
z_1^{t-1}, \mathbf x_1^{t-1}, z_t )$ is $\mu(\mathbf
z_1^{t-1}, \mathbf x_1^{t-1}, z_t )$.
We define the point effect of treatment $z_t>0$ on  stratum
$(\mathbf z_1^{t-1}, \mathbf x_1^{t-1})$   by
\begin{equation}\label{eq10}
\theta(\mathbf z_1^{t-1}, \mathbf x_1^{t-1}, z_t)=\mu(\mathbf
z_1^{t-1}, \mathbf x_1^{t-1}, z_t ) -\mu(\mathbf z_1^{t-1}, \mathbf
x_1^{t-1}, z_t=0)
\end{equation}
for $t=1,\ldots, T$.

The mean   of   $y$ in stratum $(\mathbf z_1^{t}, \mathbf x_1^{t})$ is
$$
\mu(\mathbf z_1^{t}, \mathbf x_1^{t})= \sum_{\mathbf z_{t+1}^{T},
\mathbf
x_{t+1}^{T-1}} \mu( \mathbf z_1^{T},\mathbf
x_1^{T-1}) {\rm pr}(\mathbf z_{t+1}^{T},\mathbf x_{t+1}^{T-1}\mid
\mathbf z_1^{t},\mathbf x_1^{t})
$$
for $t=1,\ldots,T-1$.
We define the point effect of   covariate   $\mathbf
x_{t}> \mathbf 0$ on   stratum $(\mathbf z_1^{t}, \mathbf x_1^{t-1})$  by
\begin{equation}\label{eq11}
\gamma(\mathbf z_1^{t}, \mathbf x_1^{t-1},\mathbf x_{t})=\mu(\mathbf
z_1^{t}, \mathbf x_1^{t-1}, \mathbf x_{t}  )-\mu(\mathbf z_1^{t},
\mathbf x_1^{t-1}, \mathbf x_{t}=\mathbf 0).
\end{equation}

We define
the grand mean  by
\begin{equation}\label{eq12}
\mu=\sum_{\mathbf z_1^{T},\mathbf x_1^{T-1}}\mu(\mathbf
z_1^{T},\mathbf x_1^{T-1}){\rm pr}(\mathbf z_1^{T},\mathbf
x_1^{T-1}).
\end{equation}
Given $\{
\mathbf z_{i1}^T, \mathbf{x}_{i1}^{T-1}\}_{i=1}^N$, the proportions of treatments and covariates can be treated as  constants.
Therefore
the  point effects of treatments, the point effects of covariates and the grand mean are
  linear functions of the standard parameters $\mu(\mathbf z_1^T,\mathbf x_1^{T-1})$
 and thus are parameters, called \textbf {point parameters}, of the outcome distribution (\ref{eq1}).

 From (\ref{eq10}-\ref{eq12}), we see that each point parameter can be expressed in terms of the standard parameters
$\mu(\mathbf z_1^T,\mathbf x_1^{T-1})$.
Conversely,
we show in Appendix $A2$ that each standard parameter  can be expressed in terms of the point parameters by
$$
\mu(\mathbf z_1^T,\mathbf x_1^{T-1}) =\sum_{t=1}^T \left
[\sum_{z_t^*} - \theta(\mathbf z_1^{t-1},\mathbf x_1^{t-1},
z_t^*){\rm pr}(z_t^*\mid \mathbf z_1^{t-1},\mathbf x_1^{t-1}) +
\theta(\mathbf z_1^{t-1},\mathbf x_1^{t-1}, z_t)\right ]+
$$
\begin{equation}\label{eq13}
\sum_{t=1}^{T-1} \left [\sum_{\mathbf x_{t}^*}
-\gamma(\mathbf z_1^{t},\mathbf x_1^{t-1}, \mathbf x_{t}^*){\rm
pr}(\mathbf x_{t}^*\mid \mathbf z_1^{t},\mathbf x_1^{t-1})+
\gamma(\mathbf z_1^{t},\mathbf x_1^{t-1}, \mathbf x_{t})\right ]+
\mu.
\end{equation}
Here we take $\theta(\mathbf z_1^{t-1},\mathbf x_1^{t-1}, z_t=0)=0$ and $\gamma(\mathbf z_1^{t},\mathbf x_1^{t-1}, \mathbf x_{t}=\mathbf 0)=0$.
 Therefore the set of all point parameters, $\Psi=\{\theta(\mathbf z_1^{t-1},\mathbf x_1^{t-1}, z_t), t=1,\ldots, T; \gamma(\mathbf z_1^{t},\mathbf x_1^{t-1}, \mathbf x_{t}), t=1,\ldots, T-1; \mu\}$,   forms  a new parametrization, called \textbf {point parametrization},  for  (\ref{eq1}).

\subsection{Pattern of net  effects of treatments versus constraint on point effects of treatments}

Combining   (\ref{eq10}) with (\ref{eq2}),  we obtain
$$
\theta(\mathbf z_1^{t-1}, \mathbf x_1^{t-1}, z_t)
 =\phi(\mathbf
z_1^{t-1},\mathbf{x}_1^{t-1},z_t)+
$$
$$
\sum_{s=t+1}^T\sum_{
\mathbf z_{t+1}^{s-1},\mathbf x_{t}^{s-1}}\sum_{z_s>0  }\phi(  \mathbf{z}_1^{s-1}, \mathbf
x_1^{s-1}, z_s){\rm pr}(
\mathbf z_{t+1}^{s-1},\mathbf x_{t}^{s-1},z_s\mid \mathbf
z_1^{t},\mathbf x_1^{t-1})-
$$
$$
\sum_{s=t+1}^T\sum_{
\mathbf z_{t+1}^{s-1},\mathbf x_{t}^{s-1}}\sum_{z_s>0  }\phi( \mathbf{z}_1^{t-1},z_t=0, \mathbf{z}_{t+1}^{s-1}, \mathbf
x_1^{s-1},z_s){\rm pr}(
\mathbf z_{t+1}^{s-1},\mathbf x_{t}^{s-1},z_s\mid \mathbf
z_1^{t-1},z_t=0, \mathbf x_1^{t-1})
$$
for  $t=1,\ldots, T-1$ and $\theta(\mathbf z_1^{T-1}, \mathbf x_1^{T-1}, z_T)
 =\phi(\mathbf
z_1^{T-1},\mathbf{x}_1^{T-1},z_T)$. The formula
 decomposes
$\theta(\mathbf z_1^{t-1}, \mathbf x_1^{t-1}, z_t)$   into the net effects  of treatments $z_s>0$ at times $s \geq t$ in strata $(\mathbf z_1^{t-1}, \mathbf x_1^{t-1}, z_t)$ versus $(\mathbf z_1^{t-1}, \mathbf x_1^{t-1}, z_t=0)$.

Suppose that the data-generating mechanism is such that  the net effects $\phi( \mathbf
z_1^{t-1},\mathbf{x}_1^{t-1},z_t)$   follow certain pattern.  One example of such patterns is
$$
\phi( \mathbf
z_1^{t-1},\mathbf{x}_1^{t-1},z_t)=\varphi_1 z_t + \varphi_2 z_{t-1}+ \varphi_3 \mathbf x_{t-1}'
$$
 for any $( \mathbf
z_1^{t-1},\mathbf{x}_1^{t-1},z_t)$ at  $t=1,\ldots, T$, where   the parameter vector  $\varphi=(\varphi_1, \varphi_2, \varphi_3)$ indexes   all the net effects.
Generally,  we consider  a pattern of  net effects described   by  a function
\begin{equation}\label{eq21}
\phi(\mathbf
z_1^{t-1},\mathbf{x}_1^{t-1},z_t)=\phi(  \mathbf
z_1^{t-1},\mathbf{x}_1^{t-1},z_t; \varphi),
\end{equation}
where the  $k$-dimensional parameter vector $\varphi=(\varphi_1,\ldots, \varphi_k)$ indexes all the net effects. We   call $\varphi$ the net effect vector.
Because the  net effects  describe the conditional distribution (\ref{eq1})  of  the observable outcome $\{y_i\}_{i=1}^N$ given the observable variables $\{(\mathbf z_{i1}^{T},\mathbf x_{i1}^{T-1})\}_{i=1}^N$, pattern (\ref{eq21})  is testable with observed data.

With the pattern and the above decomposition, we obtain the constraint on point effects of treatments
\begin{equation}\label{eq22}
\theta(\mathbf z_1^{t-1}, \mathbf x_1^{t-1}, z_t)
 =\phi(\mathbf
z_1^{t-1},\mathbf{x}_1^{t-1},z_t; \varphi)+
\end{equation}
$$
\sum_{s=t+1}^T\sum_{
\mathbf z_{t+1}^{s-1},\mathbf x_{t}^{s-1}}\sum_{z_s>0  }\phi(  \mathbf{z}_1^{s-1}, \mathbf
x_1^{s-1}, z_s; \varphi){\rm pr}(
\mathbf z_{t+1}^{s-1},\mathbf x_{t}^{s-1},z_s\mid \mathbf
z_1^{t},\mathbf x_1^{t-1})-
$$
$$
\sum_{s=t+1}^T\sum_{
\mathbf z_{t+1}^{s-1},\mathbf x_{t}^{s-1}}\sum_{z_s>0  }\phi( \mathbf{z}_1^{t-1},z_t=0, \mathbf{z}_{t+1}^{s-1}, \mathbf
x_1^{s-1},z_s; \varphi){\rm pr}(
\mathbf z_{t+1}^{s-1},\mathbf x_{t}^{s-1},z_s\mid \mathbf
z_1^{t-1},z_t=0, \mathbf x_1^{t-1})
$$
for  $t=1,\ldots, T-1$ and $\theta(\mathbf z_1^{T-1}, \mathbf x_1^{T-1}, z_T)
 =\phi(\mathbf
z_1^{T-1},\mathbf{x}_1^{T-1},z_T; \varphi)$.

If the function  $\phi(\mathbf
z_1^{t-1},\mathbf{x}_1^{t-1},z_t; \varphi )$ in pattern (\ref{eq21}) correctly describes  the net effect $\phi(\mathbf
z_1^{t-1},\mathbf{x}_1^{t-1},z_t)$, then constraint
 (\ref{eq22})
does not necessarily bias  the estimate of $\phi(\mathbf
z_1^{t-1},\mathbf{x}_1^{t-1},z_t)$ ($t=1,\ldots, T$).  As a result,  constraint (\ref{eq22})
  does not necessarily lead to automatic rejection of the null hypothesis of the net effects.

\section{ML Estimation of  Net Effects of Treatments  through Point Effects of Treatments}
The data set is independent observations $\{\mathbf z_{i1}^{T},
\mathbf x_{i1}^{T-1},y_i\}$ on units $i=1,\ldots,N$.
Using the   outcome distribution (\ref{eq1}), we obtain
the following likelihood of the point parameters
\begin{equation}\label{eq41}
L\{\Psi; \{{y}_i\}_{i=1}^N|\{\mathbf z_{i1}^{T}, \mathbf x_{i1}^{T-1}\}_{i=1}^N\}
=\prod_{i=1}^N f\{y_i\mid \mathbf z_{i1}^T, \mathbf{x}_{i1}^{T-1};  \mu(\mathbf z_{i1}^T,\mathbf x_{i1}^{T-1})\}
\end{equation}
where $\Psi$ is the set of  point parameters constructed in Section $3.1$ and $\mu(\mathbf z_{i1}^{T},\mathbf x_{i1}^{T-1})=\mu(\mathbf z_{1}^{T}=\mathbf z_{i1}^{T},\mathbf x_{1}^{T-1}=\mathbf x_{i1}^{T-1})$   is   expressed  by (\ref{eq13}) in terms of the point parameters in $\Psi$.
The outcome model is
\begin{equation}\label{eq42}
\mu_i=\mu(\mathbf z_{i1}^{T},\mathbf x_{i1}^{T-1})
\end{equation}
where $\mu_i=E(y_i\mid \mathbf z_{i1}^{T},\mathbf x_{i1}^{T-1})$ is the mean of $y_i$ given $(\mathbf z_{i1}^{T},\mathbf x_{i1}^{T-1})$. The constraint on the point parameters is (\ref{eq22}).

For  common distributions, the net effects of treatments can be estimated according to the following procedure.
First, we    estimate $\mu(\mathbf z_1^{t-1}, \mathbf x_1^{t-1}, z_t)$  ($t=1,\ldots, T$) by using likelihood (\ref{eq41}) and model (\ref{eq42}).
The   estimate $\hat \mu(\mathbf z_1^{t-1},\mathbf x_1^{t-1},
z_t)$  is the average of $y$ in stratum $(\mathbf z_1^{t-1},\mathbf x_1^{t-1},
z_t)$. Second, we use $\hat \mu(\mathbf z_1^{t-1},\mathbf x_1^{t-1},
z_t)$ to calculate the estimate of $\theta(\mathbf z_1^{t-1}, \mathbf x_1^{t-1}, z_t)$ according to (\ref{eq10}).
Third, we   perform a regression of $\hat\theta(\mathbf z_1^{t-1}, \mathbf x_1^{t-1}, z_t)$ on    ${\rm pr}(
\mathbf z_{t+1}^{s-1},\mathbf x_{t}^{s-1},z_s\mid \mathbf
z_1^{t},\mathbf x_1^{t-1}) $ and ${\rm pr}(
\mathbf z_{t+1}^{s-1},\mathbf x_{t}^{s-1},z_s\mid \mathbf
z_1^{t-1},z_t=0, \mathbf x_1^{t-1})$ according to constraint  (\ref{eq22}) to  estimate $\varphi$. Finally, we replace $\varphi$ by $\hat\varphi$   in pattern (\ref{eq21}) to  obtain the estimate of $\phi(\mathbf z_1^{t-1},\mathbf x_1^{t-1},
z_t)$, that is, $\hat\phi(\mathbf z_1^{t-1},\mathbf x_1^{t-1},
z_t)=\phi(\mathbf z_1^{t-1},\mathbf x_1^{t-1},
z_t; \hat\varphi)$.
The procedure will be further illustrated in the next section. Here we analyse unbiasedness and consistency of  $\hat\phi(\mathbf
z_1^{t-1},\mathbf{x}_1^{t-1},z_t)$.

The   estimate $\hat \mu(\mathbf z_1^{t-1},\mathbf x_1^{t-1},
z_t)$   is  unbiased and so is   $\hat\theta(\mathbf z_1^{t-1},\mathbf
x_1^{t-1},z_t)$.
If $\phi(\mathbf z_1^{t-1},\mathbf x_1^{t-1},
z_t;\varphi)$ is a linear function of $\varphi$, then the estimate $\hat\varphi$ is unbiased according to (\ref{eq22}) treated as a regression model, and so is $\hat\phi(\mathbf z_1^{t-1},\mathbf x_1^{t-1},
z_t)$.
If $\phi(\mathbf z_1^{t-1},\mathbf x_1^{t-1},
z_t;\varphi)$ is not linear in $\varphi$, then $\hat\varphi$ may be biased, but $\hat\phi(\mathbf z_1^{t-1},\mathbf x_1^{t-1},
z_t)$ is usually unbiased.

Oftentimes, the dimension $k$ of  $\varphi$ is   finite, that is,  the net effects $\phi(\mathbf z_1^{t-1},\mathbf x_1^{t-1},
z_t)$  have a pattern of finite dimension.
From (\ref{eq22}) treated as a regression model, we see that $\hat \varphi$ is consistent and so is $\hat\phi(\mathbf z_1^{t-1},\mathbf x_1^{t-1},
z_t)$ if there exist at least $k$ different point effects $\theta(\mathbf z_1^{t-1},\mathbf x_1^{t-1},z_t)$ of treatments which contain $\varphi$ and whose estimates have  zero covariance matrices as the sample size $N$ tends to infinity.
This condition can be satisfied in many practical applications, where the treatment variable $z_t$ ($t=1,\ldots, T$) and the covariate $\mathbf x_t$ ($t=1,\ldots, T-1$) take  finite numbers of values.

\section{Example: Net Effect of Bonus  on Productivity of  an Employee}
\subsection{ Backgrounds and the setting}
To improve  productivity,  an employer rewards bonuses to the employees each month.  When bonus is rewarded, consideration is given to performance of an employee in the past month.
 The employer wishes to know  how the bonus influences the  productivity of an  employee over a period of more than one month. If the  bonus remained effective on productivity after   one month, then the employer would
reward bonuses less frequently.

In this context, bonuses form  a treatment sequence. The productivity after a last bonus is the outcome of interest. The performance is a covariate  between bonuses  that is simultaneously posttreatment variable of the previous bonuses and confounder of the subsequent bonuses.  The interest of the employer is the net effect of each treatment in treatment sequence  on the outcome.
Formally, the treatment variable  $z_t$ is binary:  $z_t=1$ if bonus is rewarded and $z_t=0$ otherwise $(t = 1, \ldots, T )$. The outcome $y$ is the productivity after last treatment $z_T$.
The   covariate $x_t$  is also binary:  $x_t=1$ if the performance  is good in the past  month and $x_t=0$ otherwise $(t = 1, \ldots, T-1 )$.

Each treatment variable $z_t$ has only one active treatment $z_t=1$ and thus one net effect $\phi( \mathbf
z_1^{t-1},\mathbf{x}_1^{t-1}, z_t=1)$ of treatment and one point effect $\theta( \mathbf
z_1^{t-1},\mathbf{x}_1^{t-1}, z_t=1)$ of treatment on stratum $(\mathbf
z_1^{t-1},\mathbf{x}_1^{t-1})$.
Denote  $\phi( \mathbf
z_1^{t-1},\mathbf{x}_1^{t-1}, z_t=1)$    by  $\phi( \mathbf
z_1^{t-1},\mathbf{x}_1^{t-1})$
and    $\theta( \mathbf
z_1^{t-1},\mathbf{x}_1^{t-1}, z_t=1)$   by $\theta( \mathbf
z_1^{t-1},\mathbf{x}_1^{t-1})$ ($t=1,\ldots, T$). In particular, $\phi(z_1=1)=\phi$ and   $\theta(z_1=1)=\theta$ at $t=1$.

\subsection{Pattern of net effects of treatments and constraint on point effects of   treatments}

First we consider the case of $T=2$.
The treatment $z_1$ has one net effect $\phi$ denoted by $\varphi_1$.
The treatment $z_2$ has four net effects $\phi(
z_1, {x}_1)$ for $(z_1,x_1)=(0,0),(0,1), (1,0), (1,1)$, and suppose that the four net effects are the same and denoted by $\varphi_2$.
Then the pattern of these net effects is
$$
\left\{ \begin{array}{l}
\phi=\varphi_1, \\
\phi(z_1,{x}_1)=\varphi_2
\end{array} \right.
$$
where $(z_1,x_1)=(0,0),(0,1), (1,0), (1,1)$.

Decomposing the point effect  of $z_1=1$ into the net effects of $z_1=1$ and $z_2=1$ in strata $z_1=1$ versus $z_1=0$ and using the pattern above, we obtain the following  constraint    on   $\theta$
\begin{equation}\label{eq60}
\theta=\varphi_{1}+\varphi_{2} \{{\rm pr}(z_2=1\mid z_1=1)-{\rm pr}(z_2=1\mid z_1=0)\},
\end{equation}
where ${\rm pr}(z_2=1\mid z_1)$ is the proportion of   $z_2=1$ in stratum $z_1$.
We can also obtain the formula by
inserting
 the pattern  into constraint  (\ref{eq22}) for $t=1$ and using the equality  $\sum_{x_2}{\rm pr}(x_2, z_2=1\mid z_1)={\rm pr}(z_2=1\mid z_1)$. Noticing $\theta(z_1,x_1)=\phi(z_1,x_1)$  at   $t=T=2$ and using the pattern above,
we obtain the  following  constraint    on   $\theta(z_1,x_1)$
\begin{equation}\label{eq60_1}
\theta(z_1,x_1)=\varphi_{2},
\end{equation}
where  $(z_1,x_1)=(0,0),(0,1), (1,0), (1,1)$.

For an arbitrary $T$,  suppose that  the pattern of the net effects is
$$
\left\{ \begin{array}{l}
\phi( \mathbf
z_1^{t-1},\mathbf{x}_1^{t-1})=\varphi_1, \ t \leq  T-2\\
\phi(z_1^{T-2},\mathbf{x}_1^{T-2})=\varphi_2, \\
\phi(z_1^{T-1},\mathbf{x}_1^{T-1})=\varphi_3.
\end{array} \right.
$$
Decomposing $\theta(\mathbf z_1^{t-1},\mathbf x_1^{t-1})$ into net effects of $z_s$ ($s \geq t$) in strata $(\mathbf z_1^{t-1},\mathbf x_1^{t-1}, z_t)$ versus $(\mathbf z_1^{t-1},\mathbf x_1^{t-1}, z_t=0)$
and using the pattern above, we obtain the  following  constraint
on $\theta(\mathbf z_1^{t-1},\mathbf x_1^{t-1})$ ($t=1,\ldots, T$)
\begin{equation}\label{eq63}
\theta(\mathbf z_1^{t-1},\mathbf x_1^{t-1})
=\varphi_{1} c^{(1)}(\mathbf z_1^{t-1},\mathbf x_1^{t-1})
+\varphi_{2} c^{(2)}(\mathbf z_1^{t-1},\mathbf x_1^{t-1})+\varphi_{3} c^{(3)}(\mathbf z_1^{t-1},\mathbf x_1^{t-1}),
\end{equation}
where
$$
c^{(1)}(\mathbf z_1^{t-1},\mathbf x_1^{t-1})=
$$
$$\sum_{s=t}^{T-2}\{{\rm pr}(z_s=1\mid \mathbf z_1^{t-1},\mathbf x_1^{t-1},z_t=1)
-{\rm pr}(z_s=1\mid \mathbf z_1^{t-1},\mathbf x_1^{t-1},z_t=0) \},
$$
which is   difference between  the sums of  proportions of the employees receiving the treatments $z_s=1$   at $s=t,\ldots, T-2$  in stratum $(\mathbf z_1^{t-1},\mathbf x_1^{t-1},z_t=1)$ versus in stratum $(\mathbf z_1^{t-1},\mathbf x_1^{t-1},z_t=0)$, and
$$
c^{(2)}(\mathbf z_1^{t-1},\mathbf x_1^{t-1})=
$$
$${\rm pr}(z_{T-1}=1\mid \mathbf z_1^{t-1},\mathbf x_1^{t-1},z_t=1)
-{\rm pr}(z_{T-1}=1\mid \mathbf z_1^{t-1},\mathbf x_1^{t-1},z_t=0) ,
$$
which is  difference between  the  proportions of the employees receiving the second last treatment $z_{T-1}=1$     in stratum $(\mathbf z_1^{t-1},\mathbf x_1^{t-1},z_t=1)$ versus in stratum $(\mathbf z_1^{t-1},\mathbf x_1^{t-1},z_t=0)$, and
$$
c^{(3)}(\mathbf z_1^{t-1},\mathbf x_1^{t-1})=
$$
$${\rm pr}(z_{T}=1\mid \mathbf z_1^{t-1},\mathbf x_1^{t-1},z_t=1)
-{\rm pr}(z_{T}=1\mid \mathbf z_1^{t-1},\mathbf x_1^{t-1},z_t=0),
$$
which is  difference between  the  proportions of the employees receiving the last treatment $z_{T}=1$     in stratum $(\mathbf z_1^{t-1},\mathbf x_1^{t-1},z_t=1)$ versus in stratum $(\mathbf z_1^{t-1},\mathbf x_1^{t-1},z_t=0)$.
The constraint can also be obtained by
inserting
the above pattern  into  (\ref{eq22}) and using the equality
 $$
 \sum_{
\mathbf z_{t+1}^{s-1},\mathbf x_{t}^{s-1}}  {\rm pr}(
\mathbf z_{t+1}^{s-1},\mathbf x_{t}^{s-1},z_s=1\mid \mathbf
z_1^{t},\mathbf x_1^{t-1})={\rm pr}(
z_s=1\mid \mathbf
z_1^{t},\mathbf x_1^{t-1}).
 $$

\subsection{ML estimates of point effects of  treatments}
Suppose that $y$ is normally distributed.
For simplicity, further suppose that  the variance is known  and equal to
one
 for any given $(\mathbf z_{1}^{T},\mathbf x_{1}^{T-1})$.
Given the data  set $\{\mathbf z_{i1}^{T},
\mathbf x_{i1}^{T-1},y_i\}_{i=1}^N$,
likelihood (\ref{eq41})  becomes
$$
\prod_{i=1}^N \frac{1}{\sqrt{2\pi}}\exp\left [-\frac{1}{2} \{y_i-\mu(\mathbf z_{i1}^{T},\mathbf x_{i1}^{T-1})\}^2\right ]
$$
where $\mu(\mathbf z_{i1}^{T},\mathbf x_{i1}^{T-1})$ is expressed in terms of point parameters by (\ref{eq13}).
The outcome model we use is (\ref{eq42}), i.e.
$$
\mu_i=\mu(\mathbf z_{i1}^{T},\mathbf x_{i1}^{T-1}).
$$

Let $s(A)$ be the set of units in stratum $A$ and $n(A)$ be the number of units in   stratum $A$. Using the likelihood and the outcome model above, we obtain
 $$
 \hat
\mu(\mathbf z_1^{t-1},\mathbf x_1^{t-1}, z_t)
=\frac{\sum_{i\in s(\mathbf
z_1^{t-1},\mathbf x_1^{t-1},z_t)}y_i}{n(\mathbf
z_1^{t-1},\mathbf x_1^{t-1},z_t)},
$$
   $${\rm var}\{ \hat
\mu(\mathbf z_1^{t-1},\mathbf x_1^{t-1},z_t)\}=\frac{1}{n(\mathbf
z_1^{t-1},\mathbf x_1^{t-1},z_t) }.
$$

Using  (\ref{eq10}), we obtain
\begin{equation}\label{eq61}
\hat \theta(\mathbf
z_1^{t-1},\mathbf x_1^{t-1})=\hat \mu(\mathbf z_1^{t-1},\mathbf
x_1^{t-1}, z_t=1)-\hat \mu(\mathbf z_1^{t-1},\mathbf x_1^{t-1},
z_t=0)
\end{equation}
$$
= \frac{\sum_{i\in s(\mathbf
z_1^{t-1},\mathbf x_1^{t-1},z_t=1)}y_i}{n(\mathbf
z_1^{t-1},\mathbf x_1^{t-1},z_t=1)} -  \frac{\sum_{i\in s(\mathbf
z_1^{t-1},\mathbf x_1^{t-1},z_t=0)}y_i}{n(\mathbf
z_1^{t-1},\mathbf x_1^{t-1},z_t=0)}
$$
for $t=1,\ldots, T$; in particular, for $t=1$,
 $$
 \hat\theta=\frac{\sum_{i\in s(
z_1=1)}y_i}{n(
z_1=1)} -  \frac{\sum_{i\in s(
z_1=0)}y_i}{n(
z_1=0)}.
 $$
We also obtain
\begin{equation}\label{eq62}
{\rm var}\{\hat \theta(\mathbf
z_1^{t-1},\mathbf x_1^{t-1})\}=  {\rm var}\{ \hat
\mu(\mathbf z_1^{t-1},\mathbf x_1^{t-1},z_t=1)\}+ {\rm var}\{ \hat
\mu(\mathbf z_1^{t-1},\mathbf x_1^{t-1},z_t=0)\}
\end{equation}
$$
=\frac{1}{n(\mathbf z_1^{t-1},\mathbf x_1^{t-1},z_t=1)}+\frac{1}{n(\mathbf z_1^{t-1},\mathbf x_1^{t-1},z_t=0)}
$$
for $t=1,\ldots, T$; in particular, for $t=1$,
$$
{\rm var}(\hat \theta)=\frac{1}{n( z_1=1)}+\frac{1}{n(z_1=0)}.
 $$

In Appendix $A3$,
we prove
\begin{PP}\label{T6}
Suppose that the outcome  $y$ is   normal and has the same known variance for all  given  $(\mathbf z_1^{T}, \mathbf x_1^{T-1})$.
Then the  score function $U_{\theta(\mathbf z_1^{t-1},\mathbf
x_1^{t-1})}$
depends  only on    $\theta(\mathbf
z_1^{t-1},\mathbf x_1^{t-1})$.
Therefore    $\hat \theta(\mathbf z_1^{t-1},\mathbf x_1^{t-1})$
is   independent of the   estimates  of all other point parameters.
\end{PP}

\subsection{ML estimates of net effects  of treatments}
Following the procedure described in Section $4$, we estimate the net effects of treatments by a regression of  the obtained estimates of the point effects of treatments on the proportions of  treatments. To have insight into the regression, we
consider the case of $T=2$.
The constraint (\ref{eq60_1}) on $\theta(z_1,x_1)$ implies
$$
\hat\varphi_{2}=\frac{\sum_{(z_1,
x_1)}
\hat\theta(z_1,x_1)/{\rm var}\{\hat\theta( z_1,
x_1 )\} } { \sum_{(
z_1 , x_1 )} 1 /{{\rm var}\{\hat\theta( z_1,
x_1 )\} } }
$$
with the variance
$$
{\rm var}(\hat\varphi_{2})=\frac{1} {  \sum_{(
z_1 , x_1 )} 1/ {\rm var}\{\hat\theta( z_1,
x_1 )\} },
$$
where  $\hat\theta(z_1,x_1)$ is given by  (\ref{eq61}) for $t=2$ and ${\rm var}\{\hat\theta( z_1,
x_1 )\}$ by (\ref{eq62}) for $t=2$.

The constraint (\ref{eq60}) on $\theta$ implies
$$
\hat\varphi_{1}=\hat\theta-\hat\varphi_{2} \{{\rm pr}(z_2=1\mid z_1=1)-{\rm pr}(z_2=1\mid z_1=0)\}
$$
$$
=\hat\theta-\frac{\sum_{(z_1,
x_1)}
\hat\theta(z_1,x_1)/ {\rm var}\{\hat\theta( z_1,
x_1 )\} } { \sum_{(
z_1 , x_1 )} 1 /{ {\rm var}\{\hat\theta( z_1,
x_1 )\}} }\{{\rm pr}(z_2=1\mid z_1=1)-{\rm pr}(z_2=1\mid z_1=0)\}
$$
where $
 \hat\theta$ is given by   (\ref{eq61}) for $t=1$.

Now we calculate the variance   ${\rm var}( \hat\varphi_1)$ and the correlation ${\rm cov}(\hat\varphi_1, \hat\varphi_2)$.
The  variance ${\rm var}(\hat\theta)$ is given by (\ref{eq62}) for $t=1$.
 Because $\hat\theta(z_1,x_1)$ are independent of  $\hat\theta$ according to Proposition \ref{T6}, we see that $\hat\varphi_2$  is independent of $\hat\theta$.  Thus we obtain
$$
{\rm var}( \hat\varphi_1)= {\rm var}(\hat\theta)+{\rm var}(\hat\varphi_2)\{{\rm pr}(z_2=1\mid z_1=1)-{\rm pr}(z_2=1\mid z_1=0)\}^2
$$
$$
= {\rm var}(\hat\theta)+\frac{1} {  \sum_{(
z_1 , x_1 )} 1/ {\rm var}\{\hat\theta( z_1,
x_1 )\} }\{{\rm pr}(z_2=1\mid z_1=1)-{\rm pr}(z_2=1\mid z_1=0)\}^2
$$
and
$$
{\rm cov}(\hat\varphi_1, \hat\varphi_2)={\rm var}(\hat\varphi_2)\{{\rm pr}(z_2=1\mid z_1=1)-{\rm pr}(z_2=1\mid z_1=0)\}
$$
$$
=\frac{1} {  \sum_{(
z_1 , x_1 )} 1/ {\rm var}\{\hat\theta( z_1,
x_1 )\} }\{{\rm pr}(z_2=1\mid z_1=1)-{\rm pr}(z_2=1\mid z_1=0)\}.
$$

For an arbitrary $T$,
we treat  constraint (\ref{eq63}) as a linear regression with unequal  variances ${\rm var}\{\hat\theta(\mathbf z_1^{t-1},\mathbf x_1^{t-1})\}$. Using the standard techniques of a linear regression,
we  regress $\hat\theta(\mathbf z_1^{t-1},\mathbf x_1^{t-1})$ on $c ^{(1)}(\mathbf z_1^{t-1},\mathbf x_1^{t-1})$ and $c ^{(2)}(\mathbf z_1^{t-1},\mathbf x_1^{t-1})$ and $c ^{(3)}(\mathbf z_1^{t-1},\mathbf x_1^{t-1})$
 to estimate   $\varphi_{1}$ and $\varphi_{2}$ and $\varphi_{3}$.

In this regression, we need  ${\rm var}\{\hat\theta(\mathbf z_1^{t-1},\mathbf x_1^{t-1})\}$, which   has been calculated by using a known variance of $y$ given $(\mathbf z_1^{T}, \mathbf x_1^{T-1})$, as described in Section $5.3$. If the variance of $y$ given $(\mathbf z_1^{T}, \mathbf x_1^{T-1})$ is unknown, we   estimate it, which is possible for short treatment sequence.
Even for   treatment sequence of median length, however, it may  not be possible to estimate this variance. In this case,
we   use the model
$$
\mu_i=\mu(\mathbf z_{i1}^t, \mathbf x_{i1}^{t-1}),
$$
where $\mu_i=E(y_i\mid \mathbf z_{i1}^t,\mathbf x_{i1}^{t-1})$ and $\mu(\mathbf z_{i1}^t, \mathbf x_{i1}^{t-1})=\mu(\mathbf z_{1}^t=\mathbf z_{i1}^t, \mathbf x_{1}^{t-1}=\mathbf x_{i1}^{t-1})$, to   estimate ${\rm var}\{ \hat\mu(\mathbf z_1^{t}, \mathbf x_1^{t-1})\}$. The estimate is
$$
\widehat{\rm var}\{ \hat\mu(\mathbf z_1^{t}, \mathbf x_1^{t-1})\}=\frac{\sum_{i\in s(\mathbf z_1^{t}, \mathbf x_1^{t-1})}\{y_i-\hat
\mu(\mathbf z_1^{t}, \mathbf x_1^{t-1})\}^2}{n(\mathbf z_1^{t}, \mathbf x_1^{t-1})\{n(\mathbf z_1^{t}, \mathbf x_1^{t-1})-1\}}.
$$
With  $\widehat{\rm var}\{\hat\mu(\mathbf z_1^{t},\mathbf x_1^{t-1})\}$, we calculate the estimate of ${\rm var}\{\hat\theta(\mathbf z_1^{t-1},\mathbf x_1^{t-1})\}$ according to (\ref{eq10}) and obtain
$$
\widehat{\rm var}\{\hat\theta(\mathbf z_1^{t-1},\mathbf x_1^{t-1})\}=\widehat{\rm var}\{ \hat\mu(\mathbf z_1^{t-1}, \mathbf x_1^{t-1}, z_t=1)\}
+\widehat{\rm var}\{ \hat\mu(\mathbf z_1^{t-1}, \mathbf x_1^{t-1}, z_t=0)\}.
$$

\subsection{ML estimates of net effects  of treatments in long treatment sequence}
For long treatment sequences,
the number of possible
strata $(\mathbf z_1^{t-1},\mathbf x_1^{t-1})$  becomes huge at large $t$. With a finite sample, most of these strata do not
have both active   and  control treatments of  the   variable   $z_t$, and so  the point effect
$ \theta(\mathbf z_1^{t-1},\mathbf x_1^{t-1},z_t)$ of treatment is not estimable on them.
However, the treatment  assignment often satisfies certain condition, which can be used to reduce the number of point parameters  in estimation of net effects of treatments in long treatment sequence.

For illustration, we consider a   Markov process, in which  the
assignment of $z_t$ $(t=1,\ldots, T)$ depends only on the latest covariate and  treatment   $(z_{t-1},
x_{t-1})$, so that,
$$
{\rm pr}(\mathbf z_1^{t-2},\mathbf
x_1^{t-2}\mid z_{t-1},
x_{t-1},z_t)={\rm pr}(\mathbf z_1^{t-2},\mathbf
x_1^{t-2}\mid
z_{t-1},  x_{t-1}).
$$
Consider the following mean of  $y$ in stratum $(z_{t-1},  x_{t-1}, z_t)$
$$
\mu(z_{t-1},  x_{t-1}, z_t)= \sum_{\mathbf z_1^{t-2},\mathbf
x_1^{t-2}} \mu( \mathbf z_1^{t},\mathbf
x_1^{t-1}) {\rm prop}(\mathbf z_1^{t-2},\mathbf
x_1^{t-2}\mid
z_{t-1}, x_{t-1}, z_t)
$$
$$
=\sum_{\mathbf z_1^{t-2},\mathbf
x_1^{t-2}} \mu( \mathbf z_1^{t},\mathbf
x_1^{t-1}) {\rm prop}(\mathbf z_1^{t-2},\mathbf
x_1^{t-2}\mid
z_{t-1}, x_{t-1}).
$$
Taking average on both sides of (\ref{eq10}) with respect to ${\rm prop}(\mathbf z_1^{t-2},\mathbf
x_1^{t-2}\mid
z_{t-1}, x_{t-1})$ and then using the equality above,  we obtain
the following point effect of treatment $z_t=1$ on stratum $(z_{t-1}, x_{t-1})$
$$
\theta(z_{t-1},   x_{t-1})=\sum_{\mathbf z_1^{t-2},\mathbf
x_1^{t-2}} \theta( \mathbf z_1^{t-1},\mathbf
x_1^{t-1}) {\rm prop}(\mathbf z_1^{t-2},\mathbf
x_1^{t-2}\mid   z_{t-1},
  x_{t-1})
$$
$$
=\mu( z_{t-1}, x_{t-1}, z_t=1)-\mu(z_{t-1},  x_{t-1}, z_t=0).
$$
Stratum $( z_{t-1}, x_{t-1})$  is much larger than stratum $(\mathbf z_1^{t-1},\mathbf
x_1^{t-1})$ for large $t$ and thus has a large  probability  of  having  both   active and control values of  $z_t$. Therefore  $\theta(z_{t-1},   x_{t-1})$   is estimable.

Taking average on both sides of  constraint (\ref{eq63}) with respect to ${\rm pr}(
\mathbf z_{1}^{t-2},\mathbf x_{1}^{t-2}\mid  z_{t-1},   x_{t-1})$, we obtain the constraint on $\theta(z_{t-1},x_{t-1})$
\begin{equation}\label{eq80}
\theta(z_{t-1},x_{t-1})
=\varphi_{1} c^{(1)}(z_{t-1},x_{t-1})
+\varphi_{2} c^{(2)}(z_{t-1},x_{t-1})+\varphi_{3} c^{(3)}(z_{t-1},x_{t-1}),
\end{equation}
with
$$
c^{(1)}( z_{t-1}, x_{t-1})=\sum_{s=t}^{T-2}\{{\rm pr}(z_s=1\mid  z_{t-1}, x_{t-1},z_t=1)
-{\rm pr}(z_s=1\mid z_{t-1}, x_{t-1},z_t=0) \},
$$
$$
c^{(2)}( z_{t-1}, x_{t-1})={\rm pr}(z_{T-1}=1\mid  z_{t-1}, x_{t-1},z_t=1)
-{\rm pr}(z_{T-1}=1\mid z_{t-1}, x_{t-1},z_t=0),
$$
$$
c^{(3)}(z_{t-1}, x_{t-1})=
{\rm pr}(z_{T}=1\mid  z_{t-1}, x_{t-1},z_t=1)
-{\rm pr}(z_{T}=1\mid  z_{t-1}, x_{t-1},z_t=0).
$$
The constant $c^{(1)}(z_{t-1},x_{t-1})$  describes difference between sums  of  proportions of the employees receiving the treatments $z_s=1$ ($s=t,\ldots, T-2$)   in stratum $(z_{t-1},x_{t-1},z_t=1)$ versus in stratum $(z_1^{t-1},x_{t-1},z_t=0)$,  and similarly for $c^{(2)}(z_{t-1},x_{t-1})$ and $c^{(3)}(z_{t-1},x_{t-1})$.

The   estimate $\hat \mu(z_{t-1},x_{t-1},
z_t)$  is the average of $y$ in stratum $(z_{t-1},x_{t-1},
z_t)$. Then $\hat \theta(z_{t-1},x_{t-1}
)=\hat \mu(z_{t-1},x_{t-1},
z_t=1)-\hat \mu(z_{t-1},x_{t-1},
z_t=0)$. Applying Proposition \ref{T6} to $\hat \theta(z_{t-1},x_{t-1}
)$ expressed in terms of $\hat \theta(\mathbf z_1^{t-1},\mathbf x_1^{t-1}
)$, we see that $\hat \theta(z_{t-1},x_{t-1}
)$ is independent of the estimates of point parameters at the other times, in particular,
$$
{\rm cov}\{\hat \theta(z_{t-1}, x_{t-1}); \hat \theta(z_{s-1}, x_{s-1})\}=0, \quad t \neq s.
$$
To obtain the variance ${\rm var}\{ \hat\theta(z_{t-1}, x_{t-1})\}$,  we can use the model
$$
\mu_i=\mu(z_{i(t-1)}, x_{i(t-1)}, z_{it}),
$$
where $\mu_i=E(y_i\mid z_{i(t-1)}, x_{i(t-1)}, z_{it})$ and $\mu(z_{i(t-1)}, x_{i(t-1)}, z_{it})=\mu(z_{t-1}=z_{i(t-1)}, x_{t-1}=x_{i(t-1)}, z_t=z_{it})$, to   estimate ${\rm var}\{ \hat\mu(z_{t-1}, x_{t-1}, z_t)\}$ and then ${\rm var}\{ \hat\theta(z_{t-1}, x_{t-1})\}$. With $\hat\theta(z_{t-1}, x_{t-1})$ and $\widehat{\rm var}\{ \hat\theta(z_{t-1}, x_{t-1})\}$, we use (\ref{eq80}) as regression model  to estimate the net effects $\varphi_{1}$, $\varphi_{2}$ and $\varphi_{3}$.

\subsection{Outcomes of other distributions}
For some outcome distributions such as binomial one,   $\hat\theta(\mathbf z_1^{t-1}, \mathbf x_1^{t-1})$  at time $t$  may be correlated with estimates of point parameters at the  other times.   On the other hand,
$\hat
\mu(\mathbf z_1^{t-1},\mathbf x_1^{t-1},z_t)$ and thus $\hat \theta(\mathbf
z_1^{t-1},\mathbf x_1^{t-1})$   are highly robust to  point parameters at time $s > t$,
so that $\hat \theta(\mathbf
z_1^{t-1},\mathbf x_1^{t-1})$  at time $t$  is weakly correlated with estimates of the point parameters at the other times and the correlation may be omitted.
Therefore
we may use the  method described in Section $5.4$
to estimate $\varphi_{1}$, $\varphi_{2}$ and $\varphi_{3}$. The situation for $\hat
\mu(z_{t-1},x_{t-1},z_t)$ and  $\hat \theta(
z_{t-1},x_{t-1})$ is similar, and we may use the method described in Section $5.5$ to estimate $\varphi_{1}$, $\varphi_{2}$ and $\varphi_{3}$ for long treatment sequence.

\section{Practical Procedure of Estimating Net Effects of Treatments: a Hypothetical Study}
We consider the same economic  example of Section $5$. For illustrative  clarity, we consider the case of $T=2$, but the same procedure  can  be used for treatment sequences with $T>2$.
For $T=2$, there are two treatment variables $z_1=0,1$ and $z_2=0,1$, one covariate $x_1=0,1$ and a normal outcome $y$.
The data  is presented   in Table $1$ whereas the economic background is described in Section $5.1$.

The hypothetical economic study is extension of a well-known hypothetical medical study  (Robins, 2009). In the original study, the variability of all the variables   is suppressed in order to illustrate the various aspects of sequential causal inference  including causal directed acyclic graph,  problems with the standard parametrization, the $G$-computation algorithm formula and  estimation methods such as the marginal structural  model and the $g$-estimation   model. In our hypothetical study, we allow variability of the outcome $y$ and estimate net effects of treatments   by maximum likelihood.

The point effect of $z_1=1$ on the sample is
$$
\theta=\mu(z_1=1)-\mu(z_1=0)
$$
and the point effect of treatment $z_2=1$ on stratum $(z_1, x_1)=(0,0), (0,1), (1,0), (1,1)$ is
$$
\theta(z_1,x_1)=\mu(z_1, x_1, z_2=1)-\mu(z_1, x_1, z_2=0).
$$
We  estimate $\theta$ and $\theta(z_1,x_1)$  by the direct calculation   described in Section $5.3$ and  present the estimates in Table 1.
The estimates $\hat\theta(z_1,x_1)$  are    independent   of $\hat\theta$ according to Proposition \ref{T6}.
Clearly, they are also independent    of one another because they are based on different strata $(z_1, x_1)$.

We first suppose that there is no pattern among net effects of treatments, i.e. every net effect of treatment is different from another.
So we have five net effects, $\phi=\phi(z_1=1)$ and $\phi(z_1,x_1)=\phi(z_1,x_1, z_2=1)$ with $(z_1, x_1)=(0,0), (0,1), (1,0), (1,1)$.
Decomposing the point effects of treatments into the net effects of treatments,
we express the point effects of treatments in terms of the net effects of treatments by
$$
\theta(z_1,x_1)=\phi(z_1, x_1), \\ \\ {\mbox {\rm for } } (z_1, x_1)=(0,0), (0,1), (1,0), (1,1),
$$
$$
\theta=\phi +\phi(z_1=1,x_1=0)\rm{pr}(x_1=0,z_2=1\mid z_1=1)
$$
$$
+\phi(z_1=1,x_1=1)\rm{pr}(x_1=1,z_2=1\mid z_1=1)
$$
$$
-\phi(z_1=0,x_1=0)\rm{pr}(x_1=0,z_2=1\mid z_1=0)
$$
$$
-\phi(z_1=0,x_1=1)\rm{pr}(x_1=1, z_2=1\mid z_1=0).
$$
The proportions in the formula are given in Table 1. By linear regression of $\hat\theta$ and $\hat\theta(z_1,x_1)$ on the proportions, we obtain the estimates $\hat\phi=30$, $\hat\phi(z_1=1, x_1=1)=-20$, and  $\hat\phi(z_1=0, x_1=0)=\hat\phi(z_1=1, x_1=0)=\hat\phi(z_1=0, x_1=1)=20$,  together with   their  covariance matrix (not shown here).

Now we find pattern of the net effects in the framework of statistical modeling.
By the usual significance test, we see that   $\phi$ is different from the other net effects at a significance level of, say, $5 \%$, and so is $\phi(z_1=1, x_1=1)$. Because $\hat\phi(z_1=0, x_1=0)= \hat\phi(z_1=0, x_1=1)=\hat\phi(z_1=1, x_1=0)$,  we hypothesize the following pattern of the net effects
$$
\left\{ \begin{array}{l}
\phi=\varphi_1, \\
\phi(z_1=0, x_1=0)= \phi(z_1=0, x_1=1)=\phi(z_1=1, x_1=0)=\varphi_2, \\
\phi(z_1=1, x_1=1)=\varphi_3.
\end{array} \right.
$$
Hence
the constraint on $\theta$ and  $\theta(z_1, x_1)$ is
$$
\theta(z_1,x_1)=\varphi_2, \\ \\ {\mbox {\rm for } } (z_1, x_1)=(0,0), (0,1), (1,0),
$$
$$
\theta(z_1=1,x_1=1)=\varphi_3,
$$
$$
\theta=\varphi_1+\varphi_3\rm{pr}(x_1=1,z_2=1\mid z_1=1)
$$
$$
+\varphi_2 \left \{ \rm{pr}(x_1=0,z_2=1\mid z_1=1) - \rm{pr}(x_1=0,z_2=1\mid z_1=0)  \right .
$$
$$
\left .
- \rm{pr}(x_1=1, z_2=1\mid z_1=0) \right \}.
$$
By the linear regression of $\hat\theta$ and $\hat\theta(z_1,x_1)$ on the proportions, we  obtain   estimates of  $\varphi_1$, $\varphi_2$ and $\varphi_3$ and their covariance matrix, which are presented in Table 2.  From the table, we see (1)  $\hat\varphi_1=30$ with ${\rm var}(\hat\varphi_1)=3.17$, indicating a strong association between $z_1$ and $y$, and (2) $\hat\varphi_3=-20$ with ${\rm var}(\hat\varphi_3)=4.4$, indicating a strong negative association between $z_2$ and $y$ given $(z_1=1, x_1=1)$.

Furthermore, if no unmeasured confounders exist as can be assessed by subject knowledge in combination of sensitivity analysis, these net effects of treatments are the causal net effects of treatments. Then
the point (1) above implies that the bonus at $t=1$ remains effective on the productivity after $T=2$ and the employer perhaps should reward bonuses once in two months. Interestingly, the point (2)
implies that the second bonus has not improved productivity if  the first one has. In this case the employees perhaps have outperformed their capability for productivity.

\section{Concluding Remarks}

In this article, we have introduced   the net effect of treatment in treatment sequence as parameter for the conditional distribution of outcome given all treatments and covariates and shown that the net effect of treatment is the causal net effect of treatment under the assumption of strongly ignorable treatment assignment.
As a result, we can estimate the net effect of treatment and evaluate  its causal interpretation  in two separately steps. We have studied estimation of the net effect of treatment whereas the causal identification can be carried out by using subject knowledge in combination of usual sensitivity analyses.
 With point parametrization and without the treatment assignment assumption, we are able to estimate the net effect of treatment by maximum likelihood in a straightforward way.

In our approach, we express pattern of net effects of treatments  by constraint on point effects of treatments. Point effects of treatments   are  the effects of  single-point treatments, so we can estimate them by standard methods. With estimates of point effects of treatments, we  estimate  net effects of treatments by treating constraint on point effects of treatments as a regression model.

Given data, model and the likelihood,  our estimates of net effects of treatments are most efficient due to the nature of maximum likelihood estimation. They are also unbiased.
Furthermore, they are consistent in many practical situations, where net effects of treatments have pattern of finite dimension.
The consistency is true even when  treatment sequence gets long and   the number of point parameters increases exponentially.
It is interesting to compare this consistency with  the inconsistency of the ML estimate of the effect of a single-point treatment in adjustment of a confounder of infinite dimension (Robins \& Ritov, 1997). In the latter case, the ML estimate of the treatment effect is highly correlated with that of the confounder of infinite dimension.

The major limitation of our approach to estimation of net effects of treatments is that the variability of treatments and covariates has been ignored.
In much of the current literature on   estimation of   causal net effects of treatments, this variability has also been ignored. No matter if the net effect of treatment is causal or not, however,
 it is important to incorporate  this variability into  the estimation. On the other hand, our method is based  on the conditional likelihood of a final outcome given all  treatments and covariates, which implies that the variability of treatments and covariates can be considered separately and based on the likelihood of treatments and covariates.

Due to the scope of this article, we  have only considered a relative simple setting: treatments are assigned at fixed times, treatments and   covariates are discrete, there is no missing data, the outcome model is linear and the point and net effects of treatments are measured by differences.
However,
  methods   are  available to estimate the effect of a single-point treatment in more complex settings. We believe that
analogous methods can be developed to estimate  net effects of treatments in treatment sequence  in more complex setting.

\section*{Appendix}

\subsection*{A1: Deriving   formula   (\ref{eq4})}
Like $z_t$, let   $z^{*}_t$ also indicate  the treatments at time $t$.
The assumption of strongly ignorable  assignment of treatment
$z^{*}_t$   ($t= 1, \ldots, T$) is
\begin{equation}\label{eq3}
\left\{ \begin{array}{l} \mathbf x_t^{T-1}(\mathbf z_t^{T-1}),
y(\mathbf z_t^{T})  \bot z^{*}_t \mid \mathbf z_1^{t-1}, \mathbf
x_1^{t-1} \\
0 < {\rm pr}(z^{*}_t\mid \mathbf z_1^{t-1},\mathbf x_1^{t-1}) < 1
\end{array} \right.
\end{equation}
for any   treatment sequence  $\mathbf z_t^T$   given the
variables $(\mathbf z_1^{t-1}, \mathbf x_1^{t-1})$.
Here
$A  \bot  B \mid C$ means that $A$ is conditionally independent of
$B$ given $C$.
The   variable $z^{*}_t$ indicates the treatments to be randomly assigned at $t$ whereas $z_t$ in
$\mathbf z_t^T$ indicates  the  treatments at $t$ in the treatment sequence. Under assumption (\ref{eq3}), we are going to derive (\ref{eq4}) from (\ref{eq2}) by mathematical induction.

From  assumption (\ref{eq3})  at $t=T$, i.e.
$$
\left\{ \begin{array}{l}
y(z_{T})  \bot z^{*}_T \mid \mathbf z_1^{T-1}, \mathbf
x_1^{T-1} \\
0 < {\rm pr}(z^{*}_T\mid \mathbf z_1^{T-1},\mathbf x_1^{T-1}) < 1,
\end{array} \right.
$$
we obtain
\begin{equation}\label{eq_a1}
\mu( \mathbf
z_1^{T-1},\mathbf{x}_1^{T-1},z_T) =E(y(z_T)\mid \mathbf
z_1^{T-1},\mathbf{x}_1^{T-1}, z_T)= E\{y( z_T)\mid
  \mathbf z_1^{T-1},\mathbf{x}_1^{T-1}\}.
\end{equation}
Combining (\ref{eq_a1}) with (\ref{eq2}) at $t=T$, we obtain
$$
\nu( \mathbf
z_1^{T-1},\mathbf{x}_1^{T-1},z_T) =\mu( \mathbf
z_1^{T-1},\mathbf{x}_1^{T-1},z_T) = E\{y( z_T)\mid
  \mathbf z_1^{T-1},\mathbf{x}_1^{T-1}\},
$$
$$
 \phi(\mathbf z_1^{T-1}, \mathbf x_1^{T-1}, z_{T})=E\{y( z_T)\mid
  \mathbf z_1^{T-1},\mathbf{x}_1^{T-1}\}-E\{y( z_T=0)\mid
  \mathbf z_1^{T-1},\mathbf{x}_1^{T-1}\}
 $$
which is (\ref{eq4}) at $t=T$.

Assuming that (\ref{eq4}) is also true  at times $T-1, \ldots, t+1$, we are going to derive (\ref{eq4})   at    $t$.
Using  (\ref{eq_a1}), we have
$$
 \mu(\mathbf z_1^{t}, \mathbf x_1^{t-1})=\sum_{\mathbf z_{t+1}^{T},
\mathbf
x_{t}^{T-1}} \mu( \mathbf z_1^{T},\mathbf
x_1^{T-1}) {\rm pr}(\mathbf z_{t+1}^{T},\mathbf x_{t}^{T-1}\mid
\mathbf z_1^{t},\mathbf x_1^{t-1})
$$
$$
= \sum_{\mathbf z_{t+1}^{T},
\mathbf
x_{t}^{T-1}}E\{y(z_T)\mid \mathbf
z_1^{T-1},\mathbf x_1^{T-1}\}{\rm pr}(\mathbf
z_{t+1}^{T},\mathbf x_{t}^{T-1}\mid \mathbf z_1^{t},\mathbf x_1^{t-1}).
$$
Let
$$
A(s)=\sum_{\mathbf z_{t+1}^{s},\mathbf x_{t}^{s-1}}
E\{y(z_s,\mathbf z_{s+1}^T=\mathbf 0)\mid \mathbf z_1^{s-1},\mathbf
x_1^{s-1}\}{\rm pr}(\mathbf z_{t+1}^{s},\mathbf x_{t}^{s-1}\mid
\mathbf z_1^{t},\mathbf x_1^{t-1})
$$
for  $s=T,\ldots, t+1$, and $A(t)=E\{y(z_t,\mathbf z_{t+1}^T=\mathbf 0)\mid \mathbf
z_1^{t-1},\mathbf x_1^{t-1}\}$.
Noticeably, $A(T)=\mu(\mathbf z_1^{t}, \mathbf
x_1^{t-1})$.

We rewrite $A(T)$ by
$$
A(T)=
$$
$$
\sum_{\mathbf z_{t+1}^{T},\mathbf x_{t}^{T-1}}[E\{y(z_T)\mid
\mathbf z_1^{T-1},\mathbf x_1^{T-1}\}-E\{y(z_T=0)\mid \mathbf
z_1^{T-1},\mathbf x_1^{T-1}\}]{\rm pr}(\mathbf
z_{t+1}^{T},\mathbf x_{t}^{T-1}\mid \mathbf z_1^{t},\mathbf x_1^{t-1})
$$
\begin{equation}\label{a3_1_0}
+\sum_{\mathbf z_{t+1}^{T},\mathbf x_{t}^{T-1}}E\{y(z_T=0)\mid
\mathbf z_1^{T-1},\mathbf x_1^{T-1}\}{\rm pr}(\mathbf z_{t+1}^{T},
\mathbf
x_{t}^{T-1}\mid \mathbf z_1^{t},\mathbf
x_1^{t-1})
\end{equation}
$$
= \sum_{\mathbf
z_{t+1}^{T-1},\mathbf x_{t}^{T-1}}\sum_{z_T>0}\phi(\mathbf z_1^{T-1}, \mathbf
x_1^{T-1},z_T){\rm pr}(\mathbf z_{t+1}^{T-1},\mathbf x_{t}^{T-1},z_T\mid
\mathbf z_1^{t},\mathbf x_1^{t-1})
$$
\begin{equation}\label{a3_1_0_1}
+\sum_{\mathbf z_{t+1}^{T-1},\mathbf x_{t}^{T-2}}E\{y(z_T=0)\mid
\mathbf z_1^{T-1},\mathbf x_1^{T-2}\}{\rm pr}(\mathbf z_{t+1}^{T-1},
\mathbf
x_{t}^{T-2}\mid \mathbf z_1^{t},\mathbf
x_1^{t-1}).
\end{equation}
Here  the first summation term in (\ref{a3_1_0})  is equal to the first summation term in  (\ref{a3_1_0_1}) according to  (\ref{eq4}) at $t=T$; the second summation term in  (\ref{a3_1_0}), after being summed up over $z_T$ and then $\mathbf x_{T-1}$, is equal  to the second summation term in (\ref{a3_1_0_1}).

Assumption (\ref{eq3}) at  $t=T-1$  implies
$$
y(z_{T-1},z_T)  \bot z^{*}_{T-1} \mid \mathbf z_1^{T-2}, \mathbf
x_1^{T-2}
$$
which implies
$$
E\{y(z_{T-1}, z_T=0)\mid \mathbf
z_1^{T-2},\mathbf x_1^{T-2}\}=E\{y(z_{T-1}, z_T=0)\mid \mathbf
z_1^{T-2},\mathbf x_1^{T-2},z_{T-1}\}
$$
$$=E\{y(z_T=0)\mid
\mathbf z_1^{T-2},\mathbf x_1^{T-2}, z_{T-1}\}.
$$
 Hence
the
second summation term in (\ref{a3_1_0_1}) is equal to
$$\sum_{\mathbf z_{t+1}^{T-1}
,\mathbf
x_{t}^{T-2}}E\{y(z_{T-1}, z_T=0)\mid \mathbf
z_1^{T-2},\mathbf x_1^{T-2}\}{\rm pr}(\mathbf
z_{t+1}^{T-1},\mathbf x_{t}^{T-2}\mid \mathbf z_1^{t},\mathbf x_1^{t-1})
$$
 which is
$A(T-1)$.

Therefore we obtain
\begin{equation}\label{a3_1_1}
A(T)=
\end{equation}
$$
\sum_{\mathbf z_{t+1}^{T-1},\mathbf x_{t}^{T-1}} \sum_{z_T>0}
\phi(\mathbf z_1^{T-1}, \mathbf x_1^{T-1},z_T) {\rm pr}(\mathbf z_{t+1}^{T-1},
\mathbf
x_{t}^{T-1},z_T\mid \mathbf z_1^{t},\mathbf
x_1^{t-1})+A(T-1).
$$
We continue with  the same procedure to rewrite  $A(T-1)$,$\cdots$,
$A(t+1)$  consecutively and obtain
$$
 \mu(\mathbf z_1^{t}, \mathbf x_1^{t-1})=
\sum_{s=t+1}^T\sum_{
\mathbf z_{t+1}^{s-1},\mathbf x_{t}^{s-1}}\sum_{z_s>0  }\phi(\mathbf z_1^{s-1}, \mathbf
x_1^{s-1},z_s){\rm pr}(
\mathbf z_{t+1}^{s-1},\mathbf x_{t}^{s-1},z_s\mid \mathbf
z_1^{t},\mathbf x_1^{t-1}).
$$
$$
+
E\{y(z_t,\mathbf z_{t+1}^T=\mathbf 0)\mid \mathbf
z_1^{t-1},\mathbf x_1^{t-1}\}.
$$
Combining this with (\ref{eq2}) at $t$, we obtain (\ref{eq4}) at $t$.

\subsection*{A2: Deriving   formula   (\ref{eq13})}
Using (\ref{eq10}) at $t=T$, we obtain
\begin{equation}\label{a2_1_1}
\mu(\mathbf z_1^{T},\mathbf x_1^{T-1})=\mu(\mathbf z_1^{T-1},\mathbf
x_1^{T-1}, z_{T}=0) + \theta(\mathbf z_1^{T-1},\mathbf x_1^{T-1},
z_T),
\end{equation}
where  we take $\theta(\mathbf z_1^{T-1},\mathbf x_1^{T-1},
z_T=0)=0$. Taking   average on both sides of (\ref{a2_1_1}) with respect to
${\rm pr}(z_T\mid \mathbf
z_1^{T-1},\mathbf x_1^{T-1})$, we obtain
$$\mu(\mathbf z_1^{T-1},\mathbf x_1^{T-1})=\mu(\mathbf
z_1^{T-1},\mathbf x_1^{T-1}, z_{T}=0) + \sum_{z_T^*}\theta(\mathbf
z_1^{T-1},\mathbf x_1^{T-1}, z_T^*) {\rm pr}(z_T^*\mid \mathbf
z_1^{T-1},\mathbf x_1^{T-1})$$
which  implies
$$
\mu(\mathbf
z_1^{T-1},\mathbf x_1^{T-1}, z_{T}=0)=-\sum_{z_T^*}\theta(\mathbf
z_1^{T-1},\mathbf x_1^{T-1}, z_T^*){\rm pr}(z_T^*\mid \mathbf
z_1^{T-1},\mathbf x_1^{T-1})+\mu(\mathbf z_1^{T-1},\mathbf
x_1^{T-1}).
$$
 Inserting this     into (\ref{a2_1_1}), we obtain
\begin{equation}\label{a2_1_1_1}
\mu(\mathbf z_1^{T},\mathbf x_1^{T-1})=
\end{equation}
$$\sum_{z_T^* }
-\theta(\mathbf z_1^{T-1},\mathbf x_1^{T-1}, z_T^*){\rm
pr}(z_T^*\mid \mathbf z_1^{T-1},\mathbf x_1^{T-1}) +\theta(\mathbf
z_1^{T-1},\mathbf x_1^{T-1}, z_T) + \mu(\mathbf z_1^{T-1},\mathbf
x_1^{T-1}).
$$

Using (\ref{eq11}) at $t=T-1$ and then following the above procedure, we obtain
\begin{equation}\label{a2_1_1_2}
\mu(\mathbf z_1^{T-1},\mathbf x_1^{T-1})=
\end{equation}
$$\sum_{\mathbf x_{T-1}^* }
-\gamma(\mathbf z_1^{T-1},\mathbf x_1^{T-2}, \mathbf x_{T-1}^*){\rm
pr}(\mathbf x_{T-1}^*\mid \mathbf z_1^{T-1},\mathbf x_1^{T-2})
+\gamma(\mathbf z_1^{T-1},\mathbf x_1^{T-2}, \mathbf x_{T-1}) +
\mu(\mathbf z_1^{T-1},\mathbf x_1^{T-2}).
$$

Inserting (\ref{a2_1_1_2}) into (\ref{a2_1_1_1}), we obtain
$$
\mu(\mathbf z_1^{T},\mathbf x_1^{T-1})=\sum_{z_T^* }
-\theta(\mathbf z_1^{T-1},\mathbf x_1^{T-1}, z_T^*){\rm
pr}(z_T^*\mid \mathbf z_1^{T-1},\mathbf x_1^{T-1}) +\theta(\mathbf
z_1^{T-1},\mathbf x_1^{T-1}, z_T)
$$
$$
+\sum_{\mathbf x_{T-1}^* }
-\gamma(\mathbf z_1^{T-1},\mathbf x_1^{T-2}, \mathbf x_{T-1}^*){\rm
pr}(\mathbf x_{T-1}^*\mid \mathbf z_1^{T-1},\mathbf x_1^{T-2})
+\gamma(\mathbf z_1^{T-1},\mathbf x_1^{T-2}, \mathbf x_{T-1})
$$
$$
+\mu(\mathbf z_1^{T-1},\mathbf x_1^{T-2}).
$$
We go on with the same procedure for $\mu(\mathbf z_1^{T-1},\mathbf
x_1^{T-2}),\ldots,\mu(z_1)$ consecutively and  finally  obtain
(\ref{eq13}).

Formula (\ref{a2_1_1_1}) is true for any $T$. Taking $T=t$,  we obtain
\begin{equation}\label{a2_1_1_3_0}
\mu(\mathbf z_1^{t},\mathbf x_1^{t-1}) =
\end{equation}
$$
 \sum_{z_t^{*}>0} -
\theta(\mathbf z_1^{t-1},\mathbf x_1^{t-1}, z_t^{*}){\rm pr}(z_t^{*}\mid
\mathbf z_1^{t-1},\mathbf x_1^{t-1}) + \theta(\mathbf
z_1^{t-1},\mathbf x_1^{t-1}, z_t)
+ \mu(\mathbf z_1^{t-1},\mathbf
x_1^{t-1})
$$
which will be used in Appendix $A3$.

\subsection*{A3: Proving Proposition  \ref{T6} }
In the  example of Section $6$, treatment $z_t$ takes either one or zero and $\theta(\mathbf
z_1^{t-1},\mathbf x_1^{t-1},z_t=1)$ is denoted by $\theta(\mathbf
z_1^{t-1},\mathbf x_1^{t-1})$.
According to
the chain rule, the score function for $\theta(\mathbf z_1^{t-1},\mathbf x_1^{t-1})$ is equal to
\begin{equation}\label{a4_0}
U_{\theta(\mathbf z_1^{t-1},\mathbf x_1^{t-1})}=\sum_{\mathbf
z_1^{*T},\mathbf x_1^{*(T-1)}}U_{\mu(\mathbf z_1^{*T},\mathbf
x_1^{*(T-1)}) }\frac{\partial \mu(\mathbf z_1^{*T},\mathbf
x_1^{*(T-1)}) }{
\partial \theta(\mathbf z_1^{t-1},\mathbf
x_1^{t-1})}.
\end{equation}

Let $I_a(x)$ be an indicator function taking one if $x=a$ and zero otherwise.
Using   formula (\ref{eq13}), which has been proved in Appendix $A2$, we obtain
$$
\mu(\mathbf z_1^{*T},\mathbf x_1^{*(T-1)}) =\sum_{t=1}^T \left
\{  - \theta(\mathbf z_1^{*(t-1)},\mathbf x_1^{*(t-1)}
){\rm pr}(z_t^{**}=1\mid \mathbf z_1^{*(t-1)},\mathbf x_1^{*(t-1)}) \right .
$$
$$
\left . +
\theta(\mathbf z_1^{*(t-1)},\mathbf x_1^{*(t-1)}) I_{1}(z_t^*) \right \}+ A
$$
where $A$ is some  function of  the terms that do not depend on  $\theta(\mathbf z_1^{*(t-1)},\mathbf x_1^{*(t-1)})$ ($t=1,\ldots,T$).
Hence we obtain
\begin{equation}\label{a4_1}
\frac{\partial
\mu(\mathbf z_1^{*T},\mathbf x_1^{*(T-1)}) }{\partial \theta(\mathbf z_1^{t-1},\mathbf
x_1^{t-1})}=
\end{equation}
$$
I_{(\mathbf z_1^{t-1},\mathbf x_1^{t-1})}(\mathbf
z_1^{*(t-1)},\mathbf x_1^{*(t-1)}) \{I_{1}(z_t^*)- {\rm
pr}(z_t^{**}=1\mid \mathbf z_1^{t-1},\mathbf x_1^{t-1}) \}.
$$
 Furthermore,
the score function $U_{\mu(\mathbf z_1^{*T},\mathbf
x_1^{*(T-1)})}$ for  $\mu(\mathbf z_1^{*T},\mathbf
x_1^{*(T-1)})$ is
\begin{equation}\label{a4_2}
 U_{\mu(\mathbf z_1^{*T},\mathbf
x_1^{*(T-1)})}=\sum_{i\in s(\mathbf z_1^{*T},\mathbf
x_1^{*(T-1)})}\{y_i- \mu(\mathbf z_1^{*T},\mathbf x_1^{*(T-1)})\}
\end{equation}
because $y$ is normal, where the variance of $y$ given $(\mathbf z_1^{*T},\mathbf x_1^{*(T-1)})$ is assumed to be one for notational simplicity.

Inserting (\ref{a4_1}) and (\ref{a4_2}) into (\ref{a4_0})
 and then summing
the expression over $(\mathbf x_1^{*(T-1)},\mathbf z_{1}^{*T} )$,
we obtain
$$
U_{\theta(\mathbf z_1^{t-1},\mathbf x_1^{t-1})}=
$$
$$
\sum_{z_t^*=0,1}
\left \{I_{1}(z_t^*)-{\rm pr}(z_t^{**}=1\mid \mathbf z_1^{t-1},\mathbf
x_1^{t-1})\right \}\sum_{i\in s(\mathbf z_1^{t-1},\mathbf x_1^{t-1},
z_t^*)}\{ y_i-\mu(\mathbf z_1^{t-1},\mathbf x_1^{t-1}, z_t^*) \}=
$$
$$
\sum_{z_t^*=0,1}
\left \{I_{1}(z_t^*)-{\rm pr}(z_t^{**}=1\mid \mathbf z_1^{t-1},\mathbf
x_1^{t-1})\right \}\left\{
\sum_{i\in s(\mathbf z_1^{t-1},\mathbf x_1^{t-1},
z_t^*)}y_i -n(\mathbf z_1^{t-1},\mathbf x_1^{t-1}, z_t^*)
\mu(\mathbf z_1^{t-1},\mathbf x_1^{t-1}, z_t^*)
\right \}
$$

Replacing $z_t$ by $z_t^*$ and $z_t^*$ by $z_t^{**}$ in formula (\ref{a2_1_1_3_0}) and noticing that $z_t^*$ and $z_t^{**}$ take either one or zero,
we obtain
$$
\mu(\mathbf z_1^{t-1},\mathbf x_1^{t-1},z_t^*) =\theta(\mathbf z_1^{t-1},\mathbf x_1^{t-1}) \left
\{ I_{1}(z_t^*)  - {\rm pr}(z_t^{**}=1\mid \mathbf z_1^{t-1},\mathbf x_1^{t-1})
\right \}+\mu(\mathbf z_1^{t-1},\mathbf x_1^{t-1}).
$$
Hence we obtain
$$
U_{\theta(\mathbf z_1^{t-1},\mathbf x_1^{t-1})}=
$$
$$
\sum_{z_t^*=0,1}
\left \{I_{1}(z_t^*)-{\rm pr}(z_t^{**}=1\mid \mathbf z_1^{t-1},\mathbf
x_1^{t-1})\right \}     [ \sum_{i\in s(\mathbf
z_1^{t-1},\mathbf x_1^{t-1}, z_t^*)} y_i  -n(\mathbf
z_1^{t-1},\mathbf x_1^{t-1},z_t^*)
$$
$$
 \theta(\mathbf z_1^{t-1},\mathbf x_1^{t-1})
\left \{ I_{1}(z_t^*)  - {\rm pr}(z_t^{**}=1\mid \mathbf z_1^{t-1},\mathbf x_1^{t-1}) \right \} -n(\mathbf z_1^{t-1},\mathbf x_1^{t-1},z_t^*)\mu(\mathbf z_1^{t-1},\mathbf x_1^{t-1})
 ].
$$

Furthermore, we  have
$$
\sum_{z_t^*=0,1}\left \{I_{1}(z_t^*)-{\rm pr}(z_t^{**}=1\mid \mathbf
z_1^{t-1},\mathbf x_1^{t-1})\right \} n(\mathbf z_1^{t-1},\mathbf
x_1^{t-1},z_t^*) \mu(\mathbf z_1^{t-1},\mathbf x_1^{t-1})=
$$
$$
\{n(\mathbf z_1^{t-1},\mathbf
x_1^{t-1},z_t^*=1)-{\rm pr}(z_t^{**}=1\mid \mathbf z_1^{t-1},\mathbf
x_1^{t-1})n(\mathbf z_1^{t-1},\mathbf
x_1^{t-1})\}\mu(\mathbf z_1^{t-1},\mathbf x_1^{t-1})=
$$
$$
\{n(\mathbf z_1^{t-1},\mathbf
x_1^{t-1},z_t^*=1)-n(\mathbf z_1^{t-1},\mathbf
x_1^{t-1},z_t^{**}=1)\}\mu(\mathbf z_1^{t-1},\mathbf x_1^{t-1})=0.
$$

Therefore we  obtain
$$
U_{\theta(\mathbf z_1^{t-1},\mathbf x_1^{t-1})}=
$$
$$
\sum_{z_t^*=0,1}
\left \{I_{1}(z_t^*)-{\rm pr}(z_t^{**}=1\mid \mathbf z_1^{t-1},\mathbf
x_1^{t-1})\right \}    [ \sum_{i\in s(\mathbf
z_1^{t-1},\mathbf x_1^{t-1}, z_t^*)} y_i  -n(\mathbf
z_1^{t-1},\mathbf x_1^{t-1},z_t^*)
$$
\begin{equation}\label{a4_3}
 \theta(\mathbf z_1^{t-1},\mathbf x_1^{t-1})
\left \{ I_{1}(z_t^*)  - {\rm pr}(z_t^{**}=1\mid \mathbf z_1^{t-1},\mathbf x_1^{t-1}) \right \}
 ].
\end{equation}
From this formula, we see that $U_{\theta(\mathbf z_1^{t-1},\mathbf x_1^{t-1})}$
depends only on   $\theta(\mathbf z_1^{t-1},\mathbf x_1^{t-1})$, thus proving the proposition.

\end{document}